\documentstyle[12pt,aaspp]{article}
\def\@journalname{Astrophysical Journal}
\def\accepted#1{\gdef\@accptdate{#1}} \accepted{\relax}
\def\journalid#1#2{\gdef\@jourvol{#1}\gdef\@jourdate{#2}}
\def\articleid#1#2{\gdef\@startpage{#1}\gdef\@finishpage{#2}}
\def\kms{{\rm\,km\,s^{-1}}}
\def\msun{{\,M_\odot}}
\def\kpc{{\rm\,kpc}}
\def\pc{{\rm\,pc}}
\def\rsun{{\,R_\odot}}
\begin{document}
\title{The vertical equilibrium of molecular gas in the galactic disk}

\author{Sangeeta Malhotra}
\affil{Princeton University Observatory, Peyton Hall, Princeton, NJ
08544 \\
 I: san@astro.princeton.edu }

\begin{abstract}

We examine the vertical structure and equilibrium of the molecular gas
layer in the galactic disk, measuring its scale height and velocity
dispersion as a function of Galactic radius by modeling the CO
emission at the tangent points.

The model takes into account emission from a large path length along
the line of sight, corresponding to an interval ($\Delta R$) of
typically a few hundred parsecs in galactic radius; and is
parametrized by the scale height of the gas, the centroid in z, the
rotation velocity and the velocity dispersion; these parameters are
assumed to be constant over the interval $\Delta R$. This model is
then fit to the $^{12}CO$ survey of Knapp et al.(1985) to determine
the best fit parameters.

The terminal velocity values are found to be in good agreement with
those obtained from HI data. The $\chi^2$ per degree of freedom from
fitting the models (considering only the photon statistics) range from
2 to 19, with a median value of 6. The main source of error is found
to be the `shot noise' due to the small number of clouds. Simulations
of the tangent point emission using discrete molecular clouds are
carried out to estimate the errors. We use the observed radial
distribution of molecular gas and the `standard' size-linewidth
relation for molecular clouds in the simulations. Modeling the
simulations gives values of $\chi^2$ per degree of freedom similar to
those obtained from modeling the observations.

The scale height of the distribution is found to increase fairly
monotonically with radius. The value of the velocity dispersion varies
from $\sim 2 \kms$ to $11 \kms$ with a typical uncertainty of $\sim 3
\kms$. The variation in velocity dispersion is consistent with a
monotonic increase with galactic radius.

The midplane mass density of the disk $\rho_0(R)$ is calculated from
the scale-height and velocity dispersion (under the assumption that the
velocity dispersion is isotropic) and is consistent with the local
value $\rho_0(R_0)$ of $0.2 \msun \pc^{-3}$ determined from
stellar kinematics.

\end{abstract}

\section{INTRODUCTION}

Because of its expected simplicity, the distribution and kinematics of
gas in the direction perpendicular to the plane of the disk of the
galaxy should have much to tell us about the forces acting on it and
its state of equilibrium. Considering the most obvious force, gravity,
and recalling that the scale height of molecular gas is much smaller
than that of stars, we expect a Gaussian distribution in z
$\rho(z)=\rho_0 \exp (-z^2/2 \sigma_z^2)$ (Spitzer 1942), if the
velocity dispersion $\sigma_v$ of the gas is constant. The Gaussian
scale height is then $\sigma_z^2=\sigma_v^2/4 \pi G \rho_0$, where
$\rho_0$ is the mid-plane mass density of the disk.

Knowing the scale height and the velocity dispersion in the vertical
direction one can estimate $\rho_0$, the mid-plane mass density of the
disk.  Studies using stellar populations (F dwarfs and K giants) as
test particles to probe the potential give estimates of both the local
disk mass density $\rho_0$ and the integral surface mass density
$\Sigma_0$ (Bahcall et al. 1992, Bahcall 1984a, Bahcall 1984b, Bahcall
1984c, Bienaym\'e et al. 1987, Kuijken \& Gilmore 1989a, Kuijken \&
Gilmore 1989b, Kuijken \& Gilmore 1989c, Kuijken \& Gilmore 1991a,
Kuijken \& Gilmore 1991b, Kuijken 1991).  In a complimentary approach
we can hope to probe the potential at different galactocentric
distances with gas (clouds) since gas emission from the whole disk can
be observed. Since the gas disk is much thinner than the stellar disk,
its vertical distribution probes only the midplane mass density. A
similar study has been carried out for HI by Celnik et. al (1979) and
Merrifield (1992).

The kinematic information about molecular clouds, especially their
velocity dispersion is important for determining their dynamics, with
each other as in collisions (influencing cloud formation/destruction,
star formation); and with the stars, increasing the velocity
dispersions of stars (Spitzer \& Schwarzschild 1951). The variation of
velocity dispersion with Galactic radius may give clues to the sources
of random kinetic energy to the clouds (Gammie et al. 1991),
star/cloud formation as a function of galactic radius etc.

Since the molecular gas forms a cold, thin layer in the disk, it may
be possible to look for signatures of disequilibrium in departures
from Gaussianity of the distribution. These departures may be due to
violent events stirring up the gas (for example supernovae), short
lifetimes of molecular clouds, infalling gas etc. or because of more
than one isothermal population with different values of velocity
dispersion $\sigma_v$.

The large scale distribution of molecular gas has been mapped using
the Massachusetts-Stony Brook galactic plane survey of $^{12}CO$ in
the plane by Sanders et al. 1986, Clemens et al. 1986; and the
Columbia survey by Cohen et al. 1986, Bronfman et al. 1989, Dame et
al. 1987. We use data from the Bell Labs $^{12}CO$ survey (Knapp et
al. 1985), consisting of strip maps in latitude because we need good
latitude coverage and sacrifice close sampling in longitude. The
details of the survey and the data as they relate to the present
analysis are presented in section 2.1.

The scale height of the molecular gas has been previously estimated by
Clemens et al. (1988) to be about 65 pc at $R=0.95 R_0$ and by Sanders
et al. (1984), Bronfman (1988), to be about 60 pc. Different surveys
thus give very consistent values of the scale height.

The same cannot be said for the velocity dispersion $\sigma_v$ of the
clouds.  The values of $\sigma_v$ found by different groups are
inconsistent by a factor of two, if not more. The velocity dispersion
for local (distance from the sun $< 3\kpc$ ) clouds has been estimated
with the additional input of their distances and the Galactic rotation
curve by Stark (1984) and Stark and Brand (1989), to be $7.8
^{+0.6}_{-0.5} \ \kms$.  Blitz et al. (1984) find $\sigma_v=5.7\pm 1.2
\ \kms$ for high latitude (hence local) clouds.  Liszt and Burton
(1983) estimate a single value of $\sigma_v=4.2 \kms$ for the inner
galaxy by measuring the dispersion of what they call `jitter', which
is the difference between the terminal velocity of HI and that of
$^{13}CO$. Clemens (1985) estimates a similar value by fitting a
Gaussian to the tangent point emission. To get rid of emission coming
from galactic radii smaller than the tangent point radius (sub-tangent
point emission) the fit is restricted to $\pm \sigma_z$ since the
emission from nearby gas extends to higher latitudes. The present
paper takes a new approach to modeling the tangent point emission by
fitting simultaneously the terminal velocity, velocity dispersion and
the scale height to the observations. We do not try to isolate the
tangent point emission but take into account emission from the
sub-tangent point gas. The modeling procedure is described in sections
2.2 and 2.3.

We use this model to get the best fit values of the aforementioned
parameters for each longitude observed. As expected the main source of
uncertainty in the parameters is due to the clumpiness of the
molecular gas distribution. To estimate these uncertainties in the
parameters measured, we carry out Monte Carlo simulations of the cloud
distribution in the galaxy, using the known radial distribution of gas
and standard cloud properties. The details of the simulations are
discussed in section 2.5.

In section 3 we present the best fit parameters, and try to interpret
their variation across the inner galaxy. The errors and systematic
shifts in these parameters are estimated from the Monte Carlo
simulations. We also estimate the midplane mass density of the disk.

Section 4 contains a discussion of different values, methods and the
definitions of the parameters characterizing the galactic distribution
as estimated by previous studies; and how consistent various
determinations are, particularly with respect to velocity dispersion
estimates. The assumptions behind this study and various caveats are
also touched upon. The conclusions of this paper are summarized in
section 5.

Appendix A contains contour maps of CO emission as a function of
latitude and velocity, for several different longitudes. Contours of
emission according to the best fitting models are superposed.

The value of the circular velocity at the Sun $\Theta_0$ is taken to
be $220 \kms$, and the assumed distance to the galactic center  is $8.5
\kpc$.

\section{METHOD}

\subsection{Data}

The observations analyzed here were made using the 7 meter
antenna at Bell Laboratories. The data and details of the
survey and the instruments are presented by Knapp et al. (1985).
Here we summarize the information necessary for the present analysis.

The survey consists of 38 strip maps (in latitude) of
the $^{12}CO(1\rightarrow 0)$ line emission. The maps are sampled at
intervals of $\Delta b= 2 \arcmin$ with a half-power beamwidth
of $100 \arcsec$. The latitude coverage is $\simeq
\pm 2\deg$, varying slightly from one line of sight to another.
The latitude extent is more than adequate for studying the tangent
point emission (except possibly for $ l \geq 60\deg$), extending
more than three scale heights both above and below the centers of
distribution in more than half the lines of sight.

The strip maps were taken at longitudes between $4\deg$ and $90\deg$,
spaced at equal intervals of $\Delta \sin l=0.025$. The velocity
resolution is $0.65 \kms$. The rms noise is typically 0.3 K but varies
slightly in each latitude strip.  Linear baselines were removed from
individual spectra.  Exceptions to this treatment are few and are
listed by Knapp et al. (1985).

We are able to analyze 22 of the 38 observed lines of sight. The
low longitude ($ 4\deg < l <17\deg$) maps were unusable due to the
lack of emission at (or reasonably near) the expected tangent point
velocities. The high longitude ($l > 61\deg$) maps were unusable
because their tangent point emission is at low velocities and is
contaminated by emission in reference positions; and because the
latitude extent of the maps is not adequate for this (almost) local
emission.

\subsection{Tangent points}

Assuming a circularly symmetric model of the galaxy, we can obtain
distances to the emission at extreme velocities in the first and
fourth quadrants. At each galactic longitude $l$, extreme velocity
emission comes from the tangent point at a galactocentric distance of
$R=R_0 \sin l$ (Figure 1). The tangent points are at a distance $d=R_0
\cos l$ from the sun.  Thus the observed angular extent $b$ gives
 the height $z=R_0 \cos l \tan b$. In the first quadrant the velocity
of the gas at radius R is given by
\begin{eqnarray}
V(R)= \frac{\Theta \sin l}{R/R_0} - \Theta_0 \sin l
\end{eqnarray}
where $\Theta$ is the circular velocity at $R$. At the tangent points
equation (1) reduces to $V_T=\Theta - \Theta_0 \sin l $.

{}From the observed $b$ extent and the terminal (extreme) velocities one
could calculate the scale height and the rotation velocity. The
velocity profiles however do not have a sharp cutoff due to the
velocity dispersion of the gas. Ideally the emission from the tangent
point T(b,v) is a bivariate Gaussian in altitude and velocity; but
emission from nearby radii is not well separated in velocity because of
the velocity dispersion of the gas.

\subsection{The velocity dispersion}

The finite velocity dispersion has two main effects that the present
analysis needs to take into account. First it makes the
velocity-to-distance conversion fuzzy, so it is difficult to separate
emission from $R$ (at the tangent point) and that from
$R^\prime (R^\prime=R+\Delta R)$, where $\Delta R$ depends on the
velocity dispersion. For cold gas (zero velocity dispersion) a
difference in velocity $\Delta V_T=7 \kms$ would correspond to
$\Delta R \simeq 270/\sin(l) \pc$ near the tangent point. Expecting
velocity dispersions of this size, it is meaningful to calculate
parameters like the scale height, rotation velocity and the velocity
dispersion, averaged over intervals $\Delta R$ of a few hundred
parsecs.

At the same time we must take into account the fact that the emission
at velocity V comes from R, and from the annulus between R and
($R+\Delta R$), so that T(b,v) is no longer a simple bivariate
Gaussian.  Celnik et al.(1979) derived the expression for the expected
line shape near terminal velocity. In this section we calculate the
2-dimensional (in latitude and velocity) profile of emission from near
the tangent point, i.e.  between $R$ and $R+\Delta R$, taking the
scale height, rotation velocity and velocity dispersion to be constant
over $\Delta R $. We expect the additional information in the apparent
latitude-extent as a function of velocity to yield more constraints on
the positions of the emitting regions.

Consider emission seen at velocities close to the terminal velocity;
($\Delta V \ll \Theta$)

\begin{equation}
V(R^{\prime})=V_T-\Delta V=\Theta-\Theta_0 \sin l-\Delta V
\end{equation}
giving
\begin{equation}
 \frac{R ^{\prime}}{R_0}=\frac{\Theta \sin l}{\Theta - \Delta V}
\end{equation}
An observer's line of sight intersects the annulus at radius
$R^{\prime}$ at two points (figure 1). The distances to the points of
intersection, $r_1$ and $r_2$ are given by
\begin{eqnarray}
r_{\scriptstyle 1 \atop \scriptstyle 2}=R_0 \cos l\mp
R^{\prime}\sqrt{1-\frac{\sin^2 l}{(R^{\prime}/R_0)^2}} \nonumber \\ & & \\
\Rightarrow r_{\scriptstyle 1 \atop \scriptstyle 2}
=R_0 \cos l \mp R_0 \sin l \sqrt{\frac{\Theta ^2}{(\Theta -\Delta V)^2}- 1}
\end{eqnarray}

For the same scale height, gas at the subtangent points will have
smaller and larger latitude extent than the tangent point gas,
corresponding to the far and the near part of the annulus at
$R^{\prime}$ $$\sigma_{z \scriptstyle 1 \atop \scriptstyle 2}(V)=\frac
{\cos l}{\cos l\,\mp \,\sin l\sqrt{2 \Delta V/\Theta}} \sigma_z(V_T)$$

Finally we must take into account velocity crowding effects. The
optical depth per velocity interval is determined by the line of sight distance
per unit velocity interval.

\begin{equation}
\frac{dr}{dv}=\frac{R_0 \Theta^2 \sin l}{{(\Theta_0 \sin l+ V(R ^{\prime}))}^3}
{\left ({ \frac{\Theta^2}{(\Theta_0 \sin l+V(R^{\prime}))^2}-1}\right)}^{-1/2}
\end{equation}

Thus the emission in the velocity interval $\Delta V$ (corresponding to
$\Delta R = \Theta_0/R_0 \sin(l)$) near the terminal
velocity, keeping parameters, $\sigma_v$, $\sigma_z$ fixed over the
velocity interval is

\begin{equation}
T(v,z)=A\sum_{1,2}\int{\frac{1}{2 \pi \sigma_V \sigma_z(V)}
\exp{\left(-\frac{(v-V)^2}{2 \sigma_V^2}-\frac{(z-z_0(V))^2}{2 \sigma_z(V)^2}
\right)}\frac{dr}{dV} dV}.
\end{equation}

\subsection{Fitting}

The model profile is calculated (eqn. 7) as a function of latitude $b$
and velocity $V$ for each longitude and a least-squares fit is made to
the data to determine the best fit parameters, namely the amplitude A,
the centroid in the vertical direction $z_0$, the vertical scale
height $\sigma_z$, the terminal velocity $V_T$ and the velocity
dispersion $\sigma_v$.

  The fitting is done by least squares minimization using the downhill
simplex routine, `amoeba' (Press et al. 1993). Contour maps of
emission as a function of latitude b and velocity V and superposed
best-fit models are given in Appendix A (Figure 12) and Figures 2 and
3a .

The velocity range over which the fit is made is determined by the
width of the extreme velocity feature. For each line of sight the
spectra at all latitudes are summed to form a composite spectrum. The
peak at the highest velocity is identified as the terminal velocity
feature. The lower velocity at which the emission drops to
half-maximum is defined as $V_{\rm half}$. We fit the emission seen at
velocities greater than $V_{\rm half}$.

To see how sensitive the parameters obtained are to a change in the
velocity range over which the fitting is done, we do the fitting for
different values of $V_{\rm half}$ for one of the longitudes $l=39$.
Figure 3a shows the various values of $V_{\rm half}$ in
relation to the terminal velocity feature.  The best fit parameters
for each of the fits is plotted against $V_{\rm half}$ in figure 3b. We
see that the parameters are fairly insensitive to the velocity range
over which the model is fit.

\subsection {Error estimation}

The above analysis holds true for a diffuse medium. For molecular
clouds we might interpret $T(b,V)$ as a smoothed probability of
finding molecular clouds at galactic latitude $b$ and velocity
$V$. The molecular gas is distributed as clouds and cloud complexes,
which sample this probability. The main uncertainty in deriving the
parameters of the models is the shot noise due to the small number of
clouds and the clumpiness of the clouds. From figures 2, 3 and 12, it
is clear that the models do not trace the detailed structure of the
gas.

If we ignore the clumpiness of the molecular gas layer and treat the
emission as if it were from a smoothly distributed medium, the
$\chi^2$ per degree of freedom is typically found to be a few
(2-19). The $\chi^2$ per degree of freedom, or reduced $\chi^2$ is
calculated by the relation
\begin{equation}
\chi^2=\frac{1}{N}\Sigma \left( \frac{T(b,V)-T_{model}}{T_{photon}}\right) ^2
\end{equation}
where $T_{photon}$ is the photon noise, and N the number of degrees of
freedom. Since the main source of noise is the clumpiness of the
medium, which is greater than the instrumental noise, we do not expect
the $\chi^2 $ to be 1. On the other hand we use the relation $\chi^2
=1 $ to estimate `realistic' noise `$T_{real}$' of the emission and
then proceed to calculate the confidence intervals on the parameters
using the rescaled noise.

While setting the $\chi^2 =1 $, we have lost information on the
goodness of fit of the models. Also `$T_{real}$' is not necessarily
normally distributed, an assumption made while estimating confidence
levels on the distribution parameters. So we carry out Monte-Carlo
simulations of the cloud population in the galaxy. This also useful
for detecting systematic errors, to the extent that we understand and
can simulate the cloud population.

If we assume the clouds to be ballistic particles, it is the locations
and velocities of the clouds that follow the distribution derived in
the last section. Clearly a large number of optically thin entities
would mimic a smooth distribution. The velocity crowding (dr/dv
$\rightarrow \infty$) at the tangent points somewhat justifies the
assumption that there is a large number of clouds. Velocity intervals
of few $\kms$ near the terminal velocity correspond to a $\kpc$ or so
of distance $r$ traversed along the line-of-sight. We would expect
$\sim 40$ clouds in the velocity interval of $7 \kms $ near the
terminal velocity (at longitude $l=35\deg$). As for `transparency',
molecular clouds show optically thick clumps and filaments with a
typical surface filling fraction of $30\%$ (e.g. Bally et
al. 1987). Thus the approximation that we are counting the number of
clouds/clumps, and that little emission from clouds is shadowed, is
perhaps not entirely unjustified.

To test the limitations of this approximation, and to get an estimate
of the confidence levels on the parameters, we simulate the cloud
population in the Galaxy, taking the total mass of the molecular gas
in the galaxy to be $2 \times 10^9 \msun$, and the proportionality
constant $X=\frac{N(H_2)}{I(CO)} {\rm \ to \ be \ } 3 \times 10^{20} cm^{-2}
K^{-1}$.

With the additional input of the size-linewidth-mass distribution,
\begin{equation}
\left(\frac{M}{100 \msun}\right)=\left( \frac{\Delta v}{0.36 \kms} \right)
^4=\left( \frac {R} {\pc} \right)^2
\end{equation}
and the mass spectrum of the clouds (Solomon \& Rivolo 1989)
\begin{equation}
\frac{dn(M)}{dM} \propto M^{-1.5}
\end{equation}
we can perform Monte-Carlo simulations of the cloud distribution in
the galaxy. The upper and lower cutoffs for the cloud mass are taken to
be $10^6 \msun$ and $10^3 \msun$ respectively. The cloud centers are
distributed in $(R,\theta,z)$. The radial distribution of the gas is
taken from earlier galactic surveys, which closely sample in longitude
(Sanders et al. 1986, Dame et al. 1987). The azimuthal distribution
at a particular radius is taken to be uniform.  In the z-direction the
clouds centers follow a Gaussian distribution with mean $z= 0 \pc$,
and scale height $\sigma_z=50 \pc$. The cloud-cloud velocity
dispersion is taken to be $\sigma_v=7 \kms$. The clouds themselves are
represented as Gaussians in z and v, with the aforementioned
size-linewidth relationship between $\sigma_z({\rm cloud})$ and
$\sigma_v({\rm cloud})$.

The clouds are then `observed' with a beam of FWHM $100\arcsec$ for 9
longitudes spanning the latitude range of the survey. The simulated
data sets have the same latitude and velocity coverage as the survey
and were modeled in the same fashion. Cloud-cloud shadowing near the
terminal velocity is considerable ($\simeq 7$ for mid-longitudes) and
is handled by scaling down the contribution from the shadowed region
of a cloud by $f=(1-0.3)^{n}$ where n is the number of clouds between
that cloud and the sun. Detailed modeling of the beam shape, cloud
sizes, radiative transfer etc. has not been carried out. A more
complicated model is perhaps not justified in view of the severe
blending of the features at the tangent point.

\section{RESULTS}

\subsection {Results of the simulations}

Examples of contour maps (in b-v) derived from the Monte-Carlo
simulations are shown in Figure 4 for longitudes $l=22.02\deg$,
$l=50.8\deg$ and $l=35.1\deg$ and the best fit models. At l=$35.1
\deg$ the terminal velocity regions are severely crowded with
typically 40 clouds in the region we are trying to fit. The
cloud-cloud shadowing goes up to 7, i.e. we might be looking `through'
up to 7 clouds near the tangent point velocities.

The parameters resulting from fitting the models to the simulations of
9 longitudes are shown in Table II and figure 5. All the longitudes
simulated have identical input distribution parameters, to test for
systematic errors as the velocity crowding changes with longitude. The
parameters measured agree well with the input parameters, with the
exception of the scale height $\sigma_z$, which is measured to be
higher than the input. It is reassuring that the $\chi^2$ per degree
of freedom for the simulated models is found to vary between 6 and 13,
similar to the values for the observations (2-19). The velocity
dispersion $\sigma_v$ has an uncertainty of $\pm 3\kms$, while the
error bars obtained from leastsquare fitting and scaled up noise
correspond to $\pm 3.5\kms$. The error-bars for $\sigma_z$ from
Monte-Carlo simulations are larger than the ones from the leastsquare
analysis. There are no systematic trends in the estimates of the
parameters with longitude.

The scale height of the gas is expected to be higher than the scale
height of the cloud centers, since $\sigma_{\rm centers}$ and
$\sigma_{\rm cloud}$ add in quadrature to give the scale height of the
gas. We repeat the simulations with clouds of different sizes at
l=$35.1 \deg$ , and the measured scale height of the gas does go down
with the size of the clouds. The values of $\sigma_z$ however are
still slightly higher than expected if the $\sigma_{\rm centers}$ and
$\sigma_{\rm cloud}$ add in quadrature. This is because of
crowding/shadowing which tends to flatten the
z-distribution. Simulations with no shadowing (i.e. transparent
clouds) and all of the mass in $10^4 \msun$ and $10^3 \msun$ clouds
reproduce the expected scale height of the gas (cf Table III). For the
standard mass-spectrum of the clouds in $10^6
\msun$ to $10^3 \msun$ clouds the contribution to scale height defined
by $\sigma_z^2({\rm cloud})= \sigma_z^2({\rm measured})-\sigma_z^2
({\rm centers})= 56 \pc$. Only one longitude in the Knapp et
al. survey shows a scale height larger than $55 \pc$ (table I), making
it difficult to accommodate the size spectrum used for the simulations.

The scale heights estimated for the Monte-Carlo simulations depend on
other properties like the density profiles of the clouds. Using clouds
with a sharp cutoff in emission at the cloud boundaries as seen in
local clouds ( e.g. Blitz \& Thaddeus 1980), the measured scale height
is lower than estimated for clouds with Gaussian density profiles
(table II). Simulations of emission at l=$35.1 \deg$ were done with
different values of scale height $\sigma_z({\rm centers})= 2, 10, 20,
30 {\rm \ and \ } 40 \pc$, and a size spectrum of $10^3-10^6
\msun$. The contribution of the cloud size to the scale height
$\sigma_z^2({\rm cloud})$ is constant $\simeq 42 \pm 2.7 \pc$ (figure
6). The scale height of molecular gas at the tangent points is
observed to be $ < 40 \pc$ along many lines of sight. If the resulting
scale height of the simulated data sets were not sensitive to cloud
properties we would conclude that this analysis favors small clouds.

Since the molecular gas is clumped, we wish to investigate the effect
of having no gas at the tangent points. It has been shown by Shane \&
Beiger-Smith (1966) that tangent points with little or no gas show a
low, broad-shouldered profile (also see figure 7). For the
distribution of clouds a lower profile may be taken to mean that
statistically there will be fewer clouds at the extreme
velocities. Since we allow the fitting program to choose the velocity
dispersion $\sigma_v$ and the terminal velocity $V_T$, missing tangent
point gas will lead to a lower value of $V_T$ and a higher value of
$\sigma_v$. The other possibility is that the fitting routine will get
confused by the small number of clouds and fit a single cloud giving a
small $\sigma_v$. The scatter in the parameters is found to be fairly
large for the simulations with no gas at the tangent points, as
expected from a smaller number of clouds (figure 8). Thus the very
high ($11 \kms$) as well as the very low (2 $\kms$) values of
$\sigma_v$ measured from the data are not believable.

\subsection{Distribution parameters}

So far we have derived the general functional form of T(b,V) and the
shape of the emission contours. We also see that the derived
parameters are robust to small changes in the fitting procedure, and
the derived uncertainties in the parameters are comparable to those
from the simulations. All the data and the best fit models are shown
in Appendix A, figure 12.

Figures 9 and 10, and table 1 show the distribution parameters $V_T,
\sigma_v, \sigma_z, z_0$ for all the longitudes. At low longitudes
(l=$17 \deg, 19 \deg$) $\sigma_v$ and $\sigma_z$ are found to have low
values but with larger error-bars, due to the relatively small amount
of CO at those radii.

The midplane position $z_0$ as derived from the fitting shows a smooth
undulation (figure 10). The scale height increases with radius R; also
the errors in $\sigma_z$ increase, mainly due to the coverage in
latitude translating to a smaller coverage for the nearby gas at
higher longitudes. A linear fit gives a gradient in scale height of
$5.5 \pc \kpc^ {-1}$.

The terminal velocity closely matches the terminal velocity curve for
HI (Gunn et al. 1979), pointing to the similarity of the HI
and CO distribution and kinematics. This also reassures us that the
tangent points are not substantially devoid of molecular gas, since HI
has been shown to be present at tangent points by the absence of
low broad-shouldered profiles (Shane and Beiger-Smith 1966).

The values of the velocity dispersion found for different lines of
sight range from $\sim 2 \kms$ to $11 \kms$. There is however a
trend for the $\sigma_v$ to increase at higher longitudes, i.e. with
increase in galactocentric radius. A straight line fit through the
$\sigma_v$ values shows an increase from 3.8 $\kms$ at 2.5 $\kpc$ to
7.1 $\kms$ at 7.5 $\kpc$.

\subsection{Midplane mass density of the disk}
Considering the condition for vertical equilibrium of the clouds in
the galactic disk, in the simplest case of a single isothermal
population of clouds, the vertical component of the gravitational force
is balanced by the pressure due to turbulent motions of the gas
$P_z=\rho_z \sigma_v^2$. The thickness of the molecular gas layer
being much less than that of the mass, the vertical component of the
force $ K_z= -4 \pi G \rho_0 z$ ($\rho_0 $ is the midplane mass
density) and the distribution of gas is a Gaussian with scale height
$\sigma_z$.  The midplane mass density of the disk is then given by

\begin{equation}
\rho_0 =\frac {\sigma_v^2}{4 \pi G \sigma_z^2}
\end{equation}

With the additional assumption of isotropy of velocity dispersion of
clouds, we can estimate $\rho_0$ from the scale height and the
velocity dispersion (Knapp 1988).

Figure 11 shows the midplane mass density of the disk computed from
equation 10 and the best fit parameters $\sigma_V$ and $\sigma_z$ for
different longitudes. We fit an exponential disk $\rho=\rho_c
\exp(-R/R_c)$, yielding a scale length $R_c$ = $5  \pc$ and
midplane mass density at the center of the disk $\rho_0(0)=1.1
\msun \pc^{-3}$. From these two values we get the midplane mass
density at sun $\rho_0(R_0) \simeq 0.2 \msun \pc^{-3}$.  The scale
lengths of $3.2 \kpc$ and $10.1 \kpc$ and central densities of
$\rho_0(0)=0.5 \msun \pc^{-3}$ and $0.07 \msun \pc^{-3}$ lie within
$66\%$ confidence limits, leading to an uncertainty of a factor $\sim
2.5$ in the value of $\rho_0(R_0)$.

The simulations however show that the parameters determined have
systematic errors. The scale height of the gas is measured to be
systematically higher than the scale height of the cloud centers. If
we knew the sizes of the molecular clouds reliably we could compensate
for that. Liszt and Burton (1981) favor a smaller (5-10 pc) typical
cloud size by comparing the morphology of the l-v diagrams of
simulations with observed data. If most of the mass of the molecular
is in small clouds the corrections to the scale height are
small. Shadowing of one cloud by another flattens the distribution and
leads to a further overestimate of the scale height. This effect
should be smaller for higher longitudes (less crowding/shadowing); so
we probably are overestimating the scale length $R_c$. Apart from these
effects seen in  the simulations, we may expect the velocity dispersion
measured to have a non-random component from streaming motions as we
are measuring the horizontal velocity dispersion. Further discussion
of this effect is postponed to the next section.

\section{DISCUSSION}
For a consistent analysis of the vertical equilibrium in the disk, we
must measure the scale heights and the velocity dispersion of the {\it
same} population. In the current analysis we do just that for the
molecular gas seen at the tangent points. The tangent point emission is
modeled in two dimensions, so the scale height and the velocity
dispersion are fit simultaneously, along with the other two free
parameters of the model, the rotation velocity and the centroid of the
z-distribution. One might worry about correlated errors in the
determination of the four quantities, for example a lower terminal
velocity estimation may lead to a higher measurement of the velocity
dispersion. While we don't know the real parameters we can estimate
the local dips and peaks in the parameters as compared to the values
smoothed over large radii. The fluctuations in different parameters do
not seem to be correlated (figures 9 and 10). Moreover the best-fit
parameters of the Monte-Carlo simulations, where we do know the actual
(i.e. input) parameters, do not show correlated errors.

We are measuring the scale height and the velocity dispersions for the
tangent point gas. The fact that we are sampling a portion of the
total gas in the inner galaxy may be of concern in that the parameters
may not be very representative. The velocity crowding at the tangent
points means that we are typically looking at $\simeq 1 \kpc$ of path
length. With the present analysis we fit the parameters to intervals
of a few hundred parsecs in Galactic radius, about $1 \kpc $ in the
line of sight distance. The typical number of clouds at tangent points
is 40 (at longitude $l=35 \deg$), so we are not finding the parameters
of just a single cloud unless the tangent point regions are severely
underpopulated.

\subsection {Tangent point analysis}

Since we expect the velocity dispersion to have values of few $\kms$,
the tangent point emission has to be modeled more carefully than if we
were seeking to derive the rotation curve. Most of the analyses of the
tangent point gas have been done to determine the rotation curve of
the gas (atomic or molecular). In this paper we are concerned more
with the determination of velocity dispersion and the scale height of
the gas. In principle the velocity dispersion and scale height could
be estimated simply by looking at the gas at forbidden velocities,
i.e. $V > V_T$. This is not very practical however for the following
two reasons. The first has to do with the calculation of the terminal
velocity; most estimates involve assumptions about the value of the
velocity dispersion, or at least its constancy (Gunn et al. 1979), or
the size-linewidth relation of the constituent clouds (Clemens 1985,
Liszt et al. 1981).  It would be quite circular then to use such a
rotation curve to calculate velocity dispersions.  One could use the
rotation curve determined from HI data to determine the velocity
dispersions for molecular gas. The disadvantage in doing so is mainly
due to the clumped nature of molecular gas, which means we see only a
few clouds at forbidden velocities. For this reason the velocity
dispersion and scale heights found by this method are very sensitive
to the terminal velocity. Shane and Bieger-Smith (1966) and Burton and
Gordon (1978) subtract the equivalent width of the terminal velocity
feature, which overestimates the velocity dispersion, and
underestimates the terminal velocity. These methods give values of
terminal velocities accurate to a few $\kms$ - about the value
expected for the velocity dispersion.

\subsection{Velocity dispersion}
Various methods have been used to estimate the velocity dispersion of
molecular gas. These are summarized by Stark \& Brand (1989). The
values of $\sigma_v$ found are also diverse, from 4.2 $\kms$ to 7.8
$\kms$.  The situation is further complicated by the definition of
velocity dispersion, and its possible confusion with streaming
motions. As a working definition we assume that any motion that
averages out for a distance of $\Delta R=200-400\pc$ does not
contribute to the velocity dispersion. This is achieved by keeping the
parameters (for example rotation velocity $\Theta$) stiff over such
intervals in radius, while fitting the terminal velocity feature
(section 2.3). In treatments where the terminal velocity is allowed to
vary between longitudes quite close to one another ( e.g. figure 3 in
Burton \& Gordon (1978)) smoothing the rotation curve to $\Delta
R=200-400\pc$ and calculating the scatter of $V_T$ around the smoothed
value would be a method equivalent to ours.

Stark (1984) and Stark and Brand (1989) estimate the velocity
dispersion of a population of local clouds, at distances of up to $3
\kpc$ from the sun. They find a velocity dispersion
$\sigma_v=7.8 ^{+0.6}_{-0.5} \kms$. Given that they do not distinguish
between streaming and random motions, and have an operational
definition similar to ours; their value of $\sigma_v$ found in the
solar neighborhood is consistent with our value.

The $\sigma_v$ determination likely to be most free of the vagaries of
the rotation curve and large scale motions is the velocity dispersion
derived for high latitude clouds by Blitz et. al. (1984) who find
$\sigma_v=5.7 \pm 1.2 \kms$ for a sample of 28 clouds, after excluding
a `pathological cloud' with velocity $-24.6\kms$. This is however a
measure of the velocity dispersion (mostly) in the vertical direction
which may be different from the line-of-sight velocity dispersion
estimated in this paper. Note that these are smaller clouds, and if
they have a velocity dispersion greater than the galactic plane GMCs
(Stark 1983) the vertical velocity dispersion value may be still
smaller than the velocity dispersion in the plane.

\subsection {Vertical distribution of the gas}

The vertical distribution of the gas is found in this study is
consistent with previous values. The scale height is found to
increase with radius. A linear fit gives a gradient in scale height of
$5.5 \pc \kpc^ {-1}$, and extrapolates to give a scale height $58
\pc$ at $\rsun$.  Similar values are found by Dame and Thaddeus
(1985), Dame et. al.  (1987), Sanders et. al. (1984), Grabelsky
et. al. (1987), and Clemens et. al. (1988). We find from the Monte
Carlo simulations that the (cloud-cloud) scale height is overestimated
if the cloud sizes are comparable to the scale height, and because of
cloud-cloud shadowing at tangent points. The second effect is expected
to be smaller at tangent points seen at higher longitude. This
introduces an error in the gradient of the scale height with
radius. The first effect (cloud size) will lead to an incorrect
constant term in the scale height-radius relation. Extrapolating the
$\sigma_z$ to $R_0$ we may under or overestimate $\sigma_z(R_0)$
depending on which of these effects dominates. Local determinations of
the scale height at $R_0$ for clouds/HII regions (Fich \& Blitz 1984)
give a value of $88 \pc$. The distinction between the distribution of
cloud centers and the scale height of the gas may be best made for
local clouds.

The positions of the CO layer centroid, $z_0$, are also consistent
with the above mentioned studies. The centroid $z_0$ also shows a dip,
going as far as $ \simeq 50 \pc$ (comparable to $\sigma_z$) south of
the midplane. Simulations made with the distribution of gas centered
at z=0 do not show large deviations from z=0 (figure 5, table II).
This feature is not seen in the studies of the vertical distribution
of HI (Celnik et al. 1979). Unfortunately it is difficult to see the
azimuthal structure of this waviness, since we can only look at the
tangent points. At any rate this smooth variation argues against
massively incorrect velocity-to-distance transforms, since that would
disturb the smoothness of such a feature if it was there, and would
require a conspiracy to produce one if it were not.

\subsection {Mid-plane mass density of the disk}

The mid-plane mass density of the disk has been derived making
numerous assumptions. We assume that the molecular gas layer is in
equilibrium, at least on the larger scales. We also assume that there
is a single population of molecular clouds at least with respect to
their velocity dispersions. There may be more than one population of
molecular clouds with different velocity dispersions. The vertical
distribution of molecular gas in that case will not be a single
Gaussian but a sum of more than one Gaussian. One way to test these
assumptions is to see whether the vertical distribution is indeed
Gaussian. We show in a subsequent paper that there exists a population
of small clouds at high z. The tail of the distribution is the
inconsistent with a Gaussian distribution. Similar high tails have
been found for HI (Lockman 1984). Given the `noise' in the data due to
the small number of molecular clouds, we do not think a more
complicated model can be derived.

We also assume that the clouds are independent ballistic particles so
the number of clouds found in any region is a Poissonian
variable. Moreover it is assumed that the distribution of cloud
centers is dictated by the dynamics of the disk alone. In fact the
occurrence of a cloud in a region of space is not a Poissonian event
as the clouds are correlated in space and organized into cloud
complexes.

To examine the vertical equilibrium the value of the {\it vertical}
velocity dispersion should be known, something possible only for local
molecular clouds.  We are {\it always} measuring the {\it azimuthal}
velocity dispersion when we look at the tangent points. The gas being
a dissipational component we expect the velocity dispersion to be
isotropic. The assumption of the isotropy of the velocity dispersion
can be checked by comparing the $\sigma_v$ derived for high latitude
clouds (in part the vertical component), which is $5.7 \pm 1.2 \kms$
(Blitz et al. 1984), with the value of the azimuthal component
(extrapolated value at the solar neighborhood), which we find to be
$\simeq 7.8 \kms $. Within the errors this value is consistent with
the velocity dispersion for the high latitude clouds and also with the
expected shape of the velocity ellipsoid.

The values of the radial ($u$), azimuthal ($v$) and the vertical ($w$)
velocity dispersion of (late type) stars in the solar neighborhood are
related as: $u \simeq2w$, $v \simeq \sqrt 2w$ (Mihalas \& Binney
1981). Given the uncertainty in the estimated velocity dispersion we
cannot determine if the velocity ellipsoid of the gas has that shape,
even locally. Support for the isotropy of the velocity dispersions
comes from the shape of the velocity ellipsoid of early type
stars. The O and B stars show fairly isotropic velocity dispersions
(Mihalas \& Binney 1981, p 423);
$\langle{v^2}\rangle/\langle{u^2}\rangle \simeq 1$, although the
vertical component $\langle{w^2}\rangle$ is consistently but slightly
lower. These stars are young enough that they show mainly the
kinematics of the interstellar medium.

We would like to check the mass density profile estimated in this
study against previous measurements; two such check points are the
scale length on the exponential disk (of both  the light and mass
distribution), and the mass density in the solar neighborhood. Using
the values of velocity dispersion and the scale height extrapolated to
solar radius we get a $\rho_0(R_0)=0.34 \msun \pc^{-3}$; extrapolation
of the exponential profile fitted to individual $\rho_0(R)$ gives
$\rho_0(R_0)=0.2 \msun \pc^{-3}$ with a formal uncertainty of a factor
of 2.5. Using the local value of vertical velocity dispersion
$\sigma_v$ (Blitz et al. 1984) gives $\rho_0(R_0)=0.18 \msun
\pc^{-3}$. These results are consistent with the local mass density
determinations from stellar kinematics $\rho_0(R_0)=0.2$ (Bahcall et
al. 1992, Bahcall 1984a, Bahcall 1984b, Bahcall 1984c) and
$\rho_0(R_0)=0.1$ (Bienaym\'e et al. 1987, Kuijken
\& Gilmore 1989a, Kuijken \& Gilmore 1989b, Kuijken \& Gilmore 1989c,
Kuijken \& Gilmore 1991a, Kuijken \& Gilmore 1991b, Kuijken 1991). The
best fit exponential mass distribution gives a scale length $5 \kpc $,
again with an uncertainty of a factor of 2. The scale length of the
light in the galactic disk has been found to be 3.4$\kpc$ (de
Vaucouleurs \& Pence 1978) and 5 $\kpc$ (van der Kruit 1987). Both the
scale length and the local $\rho_0(R_0)$ found in the present study
are consistent with the values of $5 \kpc$ and $0.1 \msun \pc^{-3}$
estimated by Merrifield (1992).

The systematic errors in this study as we know them are as
follows. The vertical velocity dispersion may be lower than the
azimuthal velocity dispersion used by a factor of $\sqrt 2$. This
would lower the mass density estimate from the one given here by a
factor of two. The scale height may be overestimated here because of
cloud size and also because of shadowing of clouds by one
another. This could raise the mass density estimate from the one
calculated here by a factor of at least 2.8. Also it may lead to an
overestimate of the scale length of the mass distribution since the
shadowing is lower at high longitudes. Given the systematic and
statistical errors none of the parameters of the mass distribution
mentioned in the last paragraph is favored over the other

\section{CONCLUSIONS}

In this work we have modeled the emission from molecular gas at the
tangent points in the first quadrant of the Galaxy and measured the
terminal velocity, the line of sight velocity dispersion, the scale
height and the deviation from the galactic plane. The modeling takes
into account emission from a large path along the line of sight to the
tangent point. The errors in the parameters are estimated by applying
the modeling technique to simulations of the galactic molecular cloud
population. We find

(1)The terminal velocities are in good agreement with the
terminal velocities from HI data, indicating that there is molecular
gas at most terminal points, and once again pointing to the similarity
of the kinematics of atomic and molecular gas.

(2)The velocity dispersion of the molecular gas increases with radius
between R$=2.5 \kpc$ and $7.5 \kpc$. A straight line fit extrapolates
to give a local value of $7.8 \kms$.

(3)The scale height also increases with the galactic radius. Again a
linear fit to the scale-height as a function of radius gives a local
value of $57 \pc$.

(4)The centroid of the a dip below the galactic plane with a maximum
excursion of $-51 \pc$ at $R/R_0 = 0.7$.

(5)The simulations of the distribution of molecular clouds with
`standard' properties show morphology similar to the data. Using these
simulations to estimate the errors in the parameters, we see that the
scale height of the gas is measured to be higher than the scale height
of the cloud centers, depending on the size of the clouds. This effect
cannot be corrected for without better knowledge of cloud properties.
The simulations also show that missing gas at the tangent points leads
to an overestimate of the velocity dispersion.

(6)The midplane mass density of the disk shows a decline with the
Galactic radius. An exponential disk model fitted to the data gives a
scale length of $5 \kpc$. The local midplane mass density obtained
from extrapolating this model is $\rho_0(R_0)=0.2 \msun
\pc^{-3}$. Both the scale length and the local mass density estimates
are uncertain to a factor of two at least.

\acknowledgements
It is a pleasure to thank G. R. Knapp for suggesting this problem and
for advice during all stages of this work. We would like to thank
J. Binney, R. Braun, W. B. Burton, C. Gammie and D. Spergel for
helpful discussions and R. H. Lupton for his very flexible graphics
package `SM'. We also thank the refree for comments. This work was
supported by NSF grant AST89-21700 to Princeton University.

\section{Appendix A}
This section shows the latitude-velocity maps of $^{12}CO (J: 1
\rightarrow 0)$ emission at different galactic longitudes from the
survey of Knapp et al. (1985) along with the best fit models for each
line of sight. The latitudes have been converted to the height above
the midplane at tangent points, according to the relation $z=b \ R_0 \
\cos l$, and the y axis is labeled as such. This conversion holds only
for the tangent point emission at extreme positive velocity.

\vfill
\pagebreak

\parindent=0pt
Figure 1: The geometry of the tangent point emission. S indicates the
position of the sun, R is the Galactic radius at the tangent point,
and ${\rm R}^{\prime}$ is a sub-tangent point. The models of the
tangent point emission take into account the emission from the annulus
between R and ${\rm R}^{\prime}$

Figure 2: Latitude-velocity maps of $^{12}CO$ emission at the
longitude l=22 $\deg$ (contour levels: 1.6, 2, 3, 4, 6, 8, 10, 12
K). The best fit model to the tangent point emission is shown
superposed on the data.  The model shows an abrupt cutoff at the
velocity $V_{\rm half}$, because the fitting is done only for $V >
V_{\rm half}$; $V_{\rm half}$ being the velocity at which the tangent
point emission is half its peak value (cf. Section 2.4).

Figure 3(a): Latitude-velocity maps of $^{12}CO$ emission at the
longitude l=39 $\deg$ (contour levels: 1.6, 2, 3, 4, 6, 8, 10, 12 K)
along with the best fit model. The vertical lines show the various
values of $V_{\rm half}$. The fitting is done for emission at
velocities $V > V_{\rm half}$.

Figure 3(b): The best fit parameters of the model for the longitude
$l=39$: the terminal velocity $V_T$, the velocity dispersion
$\sigma_v$, the centroid in z - $Z_0$ and the scale height $\sigma_z$
are plotted as a function of the cutoff velocity $V_{\rm
half}$. Emission at velocities $ > V_{\rm half}$ was modeled. This
figure shows that the parameters of the model are not very sensitive
to small changes in the boundaries of the region they are fit.

Figure 4(a), 4(b) and 4(c): Monte-Carlo simulations of b-v emission
maps at longitudes l=$22.02 \deg $, l=$35.1 \deg $ and l=$55.59 \deg $
respectively. The morphology may be compared to the observations at
the corresponding longitudes (Figure 12). Ten simulated b-v maps each
for these and other longitudes (l=$25 \deg, 28 \deg, 32 \deg, 39 \deg,
42\deg, 46 \deg$) were modeled to estimate the uncertainties in the
parameters. The best-fit modela are shown superposed on the simulations.
The side panels show the z-profile of the tangent point gas, summed
over V$> V_{\rm half}$, along with the best fit model. The top panels
shows the composite spectrum obtained by summing the emission over all
observed latitudes along with the model similarly summed.

Figure 5: The average and standard deviations of the best fit
parameters for the Monte-Carlo simulations of emission seen at
longitudes l=$22 \deg, 25 \deg, 28 \deg, 32 \deg, 35 \deg, 39 \deg, 42
\deg, 46 \deg$ and $51 \deg$. Ten simulations were done for each
longitude. The parameters are - the terminal velocity $V_T$, the
velocity dispersion $\sigma_v$, the centroid in z - $Z_0$ and the
scale height $\sigma_z$ and are plotted against $R/R_0= \sin({\rm
l})$. The error bars indicate the standard deviation of the measured
parameters in each sample of ten simulations.  Horizontal bars show
the input parameters. We see that the measured scale height is
systematically higher.

Figure 6: The measured scale height for various simulations for the
longitude l= $35.1 \deg$ are plotted against the input scale height
for the distribution of cloud centers (open points). Ten simulations
were done for each input scale height, and the error bars indicate the
variance of the measured scale height in each sample of ten. The
contribution due to the large sizes of molecular clouds is calculated
as $\sigma_z^2({\rm cloud})= \sigma_z^2({\rm
measured})-\sigma_z^2({\rm centers})$. The triangles show the values
of $\sigma_z^2({\rm cloud})$, which are nearly constant and have an
average of $\simeq 42 \pc$ (solid line).

Figure 7: The effect of missing tangent point gas. The lower panel
shows the velocity crowding $dr/dv$ vs the velocity at the longitude
l=$35.1 \deg$ (dotted line). The dashed curve is the convolution of the
$dr/dv$ function with a Gaussian of standard deviation 7,
corresponding to a velocity dispersion $\sigma_v=7 \kms$, and is the
expected shape of a spectrum. We get the solid curve if there is gas
missing from the tangent point. The main effect is to broaden the
profile. The upper panel shows a similar effect, looking at a
2-dimensional latitude-velocity map, which is limited in latitude. The
thick lines show a composite spectrum made by summing the emission in
latitude.

Figure 8: The effect of missing tangent point gas. The best fit
parameters for the Monte-Carlo simulations of emission seen at
longitude l=$35.1 \deg$ are seen. Triangles represent the parameters
obtained when gas is missing from the tangent points, open circles
represent the parameters for the same simulations but with the tangent
point gas present. The best-fit parameters (especially $\sigma_v$)
show more scatter when there is missing tangent point gas.

Figure 9: The velocity dispersion $\sigma_z$ and the terminal velocity
$V_T$ for the tangent point gas at different Galactic longitudes $l$
(hence Galactic radii $R/R_0= \sin(l)$). The solid line shows the
terminal velocity for HI (from Gunn et al. 1979).

Figure 10: The scale height $\sigma_z$ and the position of the centroid
of the tangent point gas $Z_0$ at different Galactic longitudes $l$
(hence Galactic radii $R/R_0=\sin(l)$). The midplane deviation from the
plane $Z_0$ at l=$48.6$ is comparable to the  scale height.

Figure 11: The midplane mass density $\rho_0(R)$ for different
Galactic radii. $\rho_0(R)$ is calculated from the scale height
$\sigma_z$ and the velocity dispersion $\sigma_v$ using equation 10
(section 3.3). The solid line shows the best fit exponential disk
model with a scale length of $5 \kpc$.

Figure 12: Latitude-velocity maps of $^{12}CO$ emission at different
galactic longitudes (central panel). The y-axis shows the height above
the plane for the tangent point gas at extreme (positive) velocities.
Superposed are the best-fit models of the tangent point emission.The
side panel shows the z-profile of the tangent point gas, summed over
V$> V_{\rm half}$, along with the best fit model. The top panel shows
the composite spectrum obtained by summing the emission over all
observed latitudes along with the model similarly summed. The models
show an abrupt cutoff at velocities $V_{\rm half}$, because the
fitting is done only for $V > V_{\rm half}$ (cf. Section 2.4). The
contour levels for the data are: 1.6, 2, 3, 4, 6, 8, 10 and 12 K. The
models are contoured at 80, 60, 40 and 30 percent of the peak
temperature.


\begin{planotable}{llrrlrl}
\tablewidth{39pc}
\tablecaption{Gas Distribution parameters}
\tablehead{
\colhead{longitude} &
\colhead{Radius} &
\colhead{Centroid}    &
\colhead{Scale height} &
\colhead{Terminal} &
\colhead{Velocity } &
\colhead{$\chi^2$} \\[.3ex]
\colhead{(in $\deg$)} &
\colhead{$(R/R_0)$} &
\colhead{$Z_0$}    &
\colhead{$\sigma_z$} &
\colhead{velocity} &
\colhead{dispersion} &
\colhead{per degree} \\[.3ex]
\colhead{} &
\colhead{$(\kpc)$} &
\colhead{$(\pc)$}    &
\colhead{$ (\pc)$} &
\colhead{ $V_T (\kms)$} &
\colhead{ $\sigma_v (\kms)$} &
\colhead{ of freedom}}

\startdata
   17.46 &$   0.300 $&$   32.44   $&$   9.97  $&$  139.0  $&$  1.8 $ &$ 1.6
$\nl
   18.97 &$   0.325 $&$   -3.11   $&$   17.96 $&$  123.9  $&$  3.6 $ &$ 2.2
$\nl
   20.49 &$   0.350 $&$   -4.94   $&$   26.26 $&$  121.7  $&$  2.7 $ &$ 3.3
$\nl
   22.02 &$   0.375 $&$    34.53  $&$   31.27 $&$  116.5  $&$  3.5 $ &$ 2.3
$\nl
   23.58 &$   0.400 $&$    17.91  $&$   43.84 $&$  105.2  $&$  6.8 $ &$
13.8$\nl
   25.25 &$   0.425 $&$    21.39  $&$   30.55 $&$  107.6  $&$  3.3 $ &$ 6.1
$\nl
   26.74 &$   0.450 $&$   -0.76   $&$   38.07 $&$  106.0  $&$  7.6 $ &$ 6.4
$\nl
   28.36 &$   0.475 $&$   -4.48   $&$   30.84  $&$ 102.9  $&$  3.4 $ &$ 4.3
$\nl
   30.0 &$   0.500 $&$   -13.78  $&$   33.56  $&$ 104.1  $&$  4.5 $ &$ 19.1$\nl
   31.67 &$   0.525 $&$    -4.10  $&$   49.87  $&$ 108.9  $&$  4.6 $ &$ 9.6
$\nl
   33.37 &$   0.550 $&$   -10.20  $&$   36.20  $&$  97.2  $&$  8.4 $ &$ 8.2
$\nl
   35.09 &$   0.575 $&$    -14.69 $&$   41.29  $&$  81.4  $&$  7.1 $ &$ 6.8
$\nl
   36.87 &$   0.600 $&$   -29.09  $&$   30.30  $&$  82.0  $&$  2.1 $ &$ 3.2$\nl
   38.68 &$   0.625 $&$    -33.96 $&$   49.05  $&$  84.4  $&$  4.2 $ &$ 5.1$\nl
   40.54 &$   0.650 $&$    -39.22 $&$   38.99  $&$  69.2  $&$  10.4$ &$ 4.9$\nl
   42.45 &$   0.675 $&$    -50.96 $&$   40.88  $&$  69.5  $&$  4.2 $ &$ 6.3$\nl
   44.43 &$   0.700 $&$   -37.47  $&$   39.59  $&$  65.5  $&$  5.9 $ &$ 5.9$\nl
   46.47 &$   0.725 $&$    -23.05 $&$   40.04  $&$  59.4  $&$  7.9 $ &$ 2.2$\nl
   48.59 &$   0.750 $&$    -19.62 $&$   52.38  $&$  55.3  $&$  4.7 $ &$ 2.9$\nl
   50.81 &$   0.775 $&$    -25.41 $&$   59.06  $&$  52.3  $&$  11.2$ &$ 8.4$\nl
   53.13 &$   0.800 $&$   -17.69  $&$   39.83  $&$  46.2  $&$  3.5 $ &$ 2.0$\nl
   55.59 &$   0.825 $&$    0.57   $&$   25.38  $&$  43.7  $&$  6.0 $ &$ 1.8$\nl
   58.21 &$   0.850 $&$    16.04  $&$   43.45  $&$  35.8  $&$  6.8 $ &$ 1.9$\nl
\end{planotable}

\begin{planotable}{lrrrrr}
\tablewidth{39pc}
\tablecaption{Monte Carlo Simulations}
\tablehead{
\colhead{Longitude} &
\colhead{Centroid}    &
\colhead{Scale height} &
\colhead{Terminal} &
\colhead{Velocity } &
\colhead{$\chi^2$} \\[.3ex]
\colhead{} &
\colhead{$Z_0$}    &
\colhead{$\sigma_z$} &
\colhead{velocity} &
\colhead{dispersion} &
\colhead{per degree} \\[.3ex]
\colhead{( in $ \deg $)} &
\colhead{$(\pc)$}    &
\colhead{$ (\pc)$} &
\colhead{ $V_T (\kms)$} &
\colhead{ $\sigma_v (\kms)$} &
\colhead{ of freedom}}

\startdata
Input &  0.0            &  50.0           & \nodata         & 7.0          &
\nodata \nl
22.02 &$ -4.7 \pm 18.0 $&$ 71.6 \pm 14.5 $&$ 112.6 \pm 5.4 $&$ 7.2 \pm 2.7$&$
6.5 $ \nl
25.25 &$ -7.7 \pm 27.0 $&$ 83.2 \pm 16.6 $&$ 108.1 \pm 3.2 $&$ 7.5 \pm 2.4$&$
11.5 $ \nl
28.36 &$  0.2 \pm 12.2 $&$ 75.3 \pm 12.4 $&$ 103.6 \pm 3.9 $&$ 5.8 \pm 2.1$&$
6.2 $ \nl
31.67 &$  7.6 \pm 9.3  $&$ 74.8 \pm 20.4 $&$ 99.5  \pm 2.9 $&$ 6.3 \pm 1.9$&$
6.9 $ \nl
35.1  &$  2.9 \pm 12.7 $&$ 74.8 \pm 26.2 $&$ 93.1  \pm 2.9 $&$ 8.3 \pm 2.9$&$
6.4 $ \nl
38.68 &$  7.4 \pm 14.6 $&$ 77.8 \pm 33.9 $&$ 77.2  \pm 5.3 $&$ 5.8 \pm 3.9$&$
12.8 $ \nl
42.45 &$ 11.2 \pm 21.3 $&$ 78.5 \pm 30.7 $&$ 70.4  \pm 3.9 $&$ 4.4 \pm 2.0$&$
10.2 $ \nl
46.47 &$ 12.1 \pm 20.9 $&$ 82   \pm 33.3 $&$ 59.1  \pm 5.4 $&$ 8.7 \pm 6.6$&$
7.6 $ \nl
50.81 &$ -1.5 \pm 15.8 $&$ 61.7 \pm 35.4 $&$ 54.7  \pm 4.8 $&$ 5.4 \pm 3.2$&$
9.4 $ \nl
average&$ 3.0 \pm 16.9 $&$ 75.5 \pm 24.8 $&   \nodata       &$ 6.6 \pm 3.1$&$
7.9 $ \nl
\end{planotable}

\begin{planotable}{lrrrrr}
\tablewidth{39pc}
\tablecaption{Monte Carlo Simulations for Different Cloud Sizes}
\tablehead{
\colhead{Mass Range} &
\colhead{Centroid}    &
\colhead{Scale height} &
\colhead{Terminal} &
\colhead{Velocity } &
\colhead{$\chi^2$} \\[.3ex]
\colhead{of the clouds} &
\colhead{$Z_0$}    &
\colhead{$\sigma_z$} &
\colhead{velocity} &
\colhead{dispersion} &
\colhead{per degree} \\[.3ex]
\colhead{$\msun$} &
\colhead{$(\pc)$}    &
\colhead{$ (\pc)$} &
\colhead{ $V_T (\kms)$} &
\colhead{ $\sigma_v (\kms)$} &
\colhead{ of freedom}}
\startdata
Input &$  0.0 $&$  50.0  $&$  93.5 $&$  7.0  $&  \nodata  \nl
$10^3-10^6$&$ 2.2\pm 12.3 $&$ 74.0 \pm 25.0 $&$ 92.8 \pm 2.9 $&$ 8.7\pm3.0 $&$
6.4 $ \nl
$10^5-10^6$&$ 3.5 \pm 6.8 $&$ 81.8 \pm 13.9 $&$ 92.9 \pm 1.6 $&$ 9.1\pm 4.2 $&$
7.6 $ \nl
$10^4-10^5$&$ 1.6\pm 6.4 $&$ 71.6 \pm 16.0 $&$ 93.4 \pm 2.3 $&$ 8.9\pm3.2 $&$
7.1 $ \nl
$10^3-10^4$&$ -1.8\pm 1.8 $&$ 57.8 \pm 5.7 $&$ 95.0 \pm 3.0 $&$ 6.3\pm2.2 $&$
6.3 $ \nl
$10^3-10^4$(no shadowing)&$ -2.0\pm 3.7 $&$ 52.2 \pm 5.5 $&$ 94.3 \pm 2.1 $&$
6.5\pm 2.1 $&$ 6.5 $ \nl
\end{planotable}

\end{document}